\documentclass[aps,10pt,twocolumn,superscriptaddress]{revtex4-1}
\usepackage{amsmath}
\usepackage{amsfonts}
\usepackage{amssymb}
\usepackage{graphicx}
\usepackage{bm}
\usepackage[colorlinks=true,linktocpage=true,breaklinks=true,allcolors=blue]{hyperref}
\usepackage{xcolor}
\begin{document}

\title{Driving spin chirality by electron dynamics in laser-excited antiferromagnets}

\author{Sumit Ghosh} % orcid id : 0000-0001-9139-7741
\email{s.ghosh@fz-juelich.de}
\affiliation{Peter Gr\"unberg Institut and Institute for Advanced Simulation 
Forschungszentrum J\"{u}lich and JARA, 52425 J\"{u}lich, Germany}

\author{Frank Freimuth} % orcid id : 0000-0001-6193-5991
\affiliation{Peter Gr\"unberg Institut and Institute for Advanced Simulation 
Forschungszentrum J\"{u}lich and JARA, 52425 J\"{u}lich, Germany}
\affiliation{Institute of Physics, Johannes Gutenberg-University Mainz, 55128 Mainz, Germany}

\author{Olena Gomonay} % orcid id : 0000-0002-9413-0337
\affiliation{Institute of Physics, Johannes Gutenberg-University Mainz, 55128 Mainz, Germany}

\author{Stefan Bl\"ugel} % orcid id : 0000-0001-9987-4733
\affiliation{Peter Gr\"unberg Institut and Institute for Advanced Simulation 
Forschungszentrum J\"{u}lich and JARA, 52425 J\"{u}lich, Germany}

\author{Yuriy Mokrousov} % orcid id : 0000-0001-6193-5991
\affiliation{Peter Gr\"unberg Institut and Institute for Advanced Simulation 
Forschungszentrum J\"{u}lich and JARA, 52425 J\"{u}lich, Germany}
\affiliation{Institute of Physics, Johannes Gutenberg-University Mainz, 55128 Mainz, Germany}

\begin{abstract}
{Optical generation of complex spin textures is one of the most exciting challenges of modern spintronics. Here, we uncover a distinct physical mechanism for imprinting spin chirality into collinear magnets with short laser pulses. By simultaneously treating the laser-ignited evolution of electronic structure and magnetic order, we show that their intertwined dynamics can result in an emergence of quasi-stable chiral states. We find that laser-driven chirality does not require any auxiliary external fields or intrinsic spin-orbit interaction to exist, and it can survive on the time scale of nanoseconds even in the presence of thermal fluctuations, which makes the uncovered mechanism relevant for understanding various optical experiments on magnetic materials. Our findings open a new perspective at the interaction of complex chiral magnetism with light.}
\end{abstract}

\maketitle

%%%%%%%%%%%%%%%%%%%%%%%%%%%%%%%%%%%%%%%%%%%
%\section{Introduction}

Chiral magnetic structures are perceived as indispensable components of the next generation of magnetic devices~\cite{Koshibae2015,Lonsky2020}, and chirality itself emerges as a robust  functional variable of magnetic systems~\cite{Menzel2012,Li2019,Balz2019,Kerber2020,Leveille2020}. The stabilisation of chiral magnetic states, such as spin-spirals~\cite{Bode2007}, multi-$q$ states~\cite{Takagi2018} or skyrmions~\cite{Sampaio2013, Jiang2015, Hirschberger2019}, has so far evolved predominantly along the lines of material and symmetry design of exchange interactions~\cite{Dupe2016,Grytsiuk2020,Paul2020}. Application of continuous external electric and magnetic fields has also emerged as a powerful tool to induce chirality by the mechanism of symmetry-breaking~\cite{Srivastava2018} or current- and field-induced interactions~\cite{ Sato2016, Freimuth2018, Karnad2018, Kato2019, Yudin2017, Stepanov2017, Lux2018, Ishihara2019, Bostrom2020}. Experimentally, localized chiral structures of various stability have been also produced by electromagnetic pulses~\cite{Romming2015}, and in particular ultra-short laser pulses \cite{Finazzi2013, Je2018, Buttner2020}. However,  theoretical understanding of underlying mechanisms which drive chiral states in ultrafast experiments is rather unsatisfactory, as it mainly relies on the effective treatment of magnetisation dynamics which does not include electronic degrees of freedom  into account explicitly~\cite{Koshibae2014, Fujita2017, Polyakov2020, Miyake2020}. While it is known that the laser-driven dynamics of electrons has a strong impact on the initial demagnetisation process~\cite{Kirilyuk2010, Krieger2015, Tows2015, Siegrist2019, Zhang2020}, it is still not clear how  electron dynamics, taking place on a much faster time scale, translates into formation of large chiral magnetic states. Recent time-dependent density functional theory studies \cite{Dewhurst2018, Hofherr2020} have successfully captured microscopic effects like optically induced spin transfer, achieving good agreement with  experimental findings, however, such studies are limited to  sub-picosecond regime and they do not give any insight into longer relaxation mechanisms ($\sim$10\,ps) \cite{Zhang2020} or formation of chiral magnetic order. Understanding  these aspects is key, since it  provides a direct link between electronic structure design and on-demand creation of macroscopic chiral magnetic objects with laser pulses.

 In this work we uncover a fundamental physical protocol for imprinting chirality in non-equilibrium spin systems. Taking one-dimensional antiferromagnetic chains as a platform, we explicitly consider the  interaction between the electric field of the laser pulse with conduction electrons, which are in turn coupled  to localized atomic spins, following the time-dependent coupled evolution of both sub-systems on equal footing. We thereby explicitly demonstrate that this intertwined dynamical process  can result in a non-thermal formation of steady chiral states. We show that chirality formation is quite robust against the thermal fluctuations which makes the uncovered mechanism relevant for various types of laser experiments performed on magnetic materials.  Despite the simple  model of the electronic structure we successfully capture the salient features of the underlying process emerging from the interplay between the electronic and magnetic interactions, thus providing a comprehensive strategy for an in-depth exploration of optically-driven chiral magnetism. 

\begin{figure}[t!]
\centering
\includegraphics[width=0.48\textwidth]{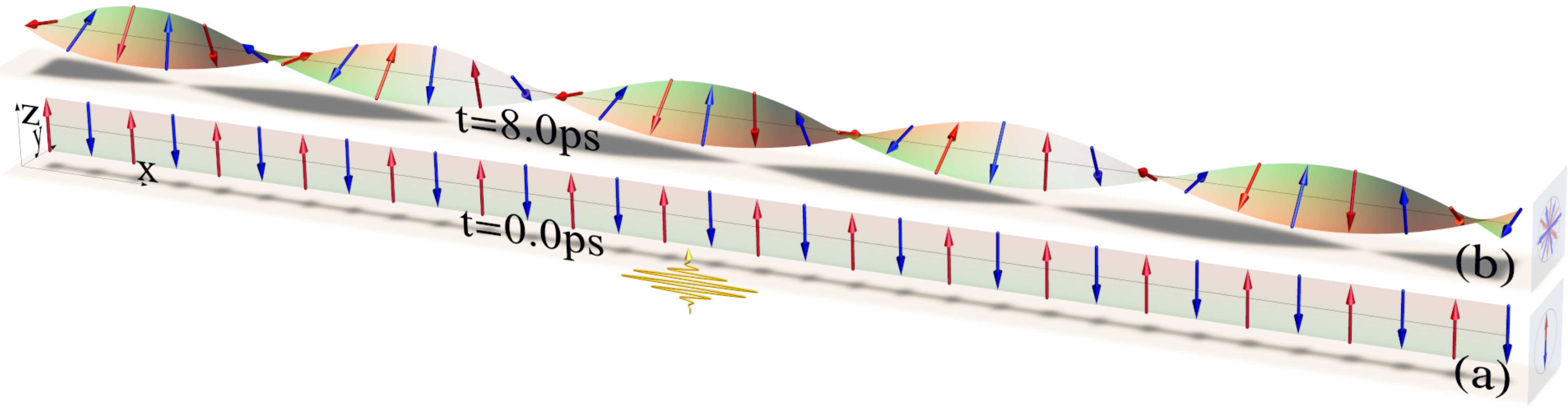}
\caption{Driving chiral antiferromagnetism by a short laser pulse polarized along the $x$-direction (yellow wavy lines). (a) Initial AFM configuration in $zx$ plane. Red and blue arrows represent initial up and down moments in the collinear AFM ground state of the system. (b) Following a complex intertwined dynamics of electrons and localized spins, a stable spin spiral state of the system is achieved \cite{SupplMov}.}
\label{fig:spiral}
\end{figure}

\section{Model and method}
Here, we focus on a simplified representative system $-$ a one-dimensional antiferromagnetic (AFM) chain of $N$ spins (see Fig.~\ref{fig:spiral}a) whose electronic properties are governed by the following double-exchange Hamiltonian~\cite{Koshibae2009, Ono2017, Ono2019}:
\begin{eqnarray}
H_0 = -J \sum_{i;\mu,\nu} c_{i,\mu}^\dagger (\hat{\mathbf{M}}_i \cdot \boldsymbol{\sigma} )_{\mu\nu} c_{i,\nu} - 
h \sum_{\langle i,j \rangle;\mu} c_{i,\mu}^\dagger  c_{j,\mu},
\label{h0} 
\end{eqnarray}
where $c_{i,\mu}$ and $c_{i,\mu}^{\dagger}$ stand for electron creation and annihilation operators at site $i$ and spin component $\mu$ and $\langle i,j \rangle$ denotes all nearest neighbour pairs. $\boldsymbol{\sigma}$ is the vector of Pauli matrices, $\hat{\mathbf{M}}_i$ is the unit vector denoting the direction of magnetic moment at site $i$. $J$  denotes the strength of coupling between the local magnetic moment and the spin moment of a conduction electron and is kept at $-$1\,eV resulting in an insulating gap of $2|J|=2$\,eV. $h$ is the hopping amplitude and is kept at 0.4\,eV. Note that the hopping takes place between the states of the same spin and thus the model does not possess any intrinsic spin flip or spin-orbit coupling. The length of the chain is fixed to $N=32$ sites. The initial ground state is constructed by filling half of the eigenstates of $H_0$ which in our model are all the states with negative eigenvalues, the Fermi level remaining at zero energy. The magnetic ground state of this linear chain is antiferromagnetic~\cite{Koshibae2009,Suppl}.
We subject the system to the action of a laser pulse, which we model as a time varying electric field in the form of a Gaussian:
\begin{eqnarray}\label{pulse}
\mathbf{\mathcal{E}}(t) &=& \mathcal{E}_0 \cos(\omega t)e^{-(t-t_0)^2/2s^2} \hat{\mathbf{x}},
\end{eqnarray}
with angular frequency $\omega = 2.02|J|/\hbar=3.07 \times 10^{15}$\,Hz  and a standard deviation $s=5$\,fs. The angular frequency is chosen to be slightly off-resonant from the bulk gap to maximise the absorption. The peak amplitude occurs at $t_0=100$\,fs and it is chosen to be $\mathcal{E}_0=0.12$\,V/$a_0$ where $a_0$ is the inter-site distance (kept at the value of 1 in our calculations). For a lattice spacing of 2.4\,$\rm \AA$ it corresponds to a field strength of 5\,MV/cm. We consider a linearly polarised pulse with polarisation along $\hat{x}$ axis and propagating along $\hat{z}$ axis (Fig.~\ref{fig:spiral}a).

\begin{figure}[t]
\centering
\includegraphics[width=0.47\textwidth]{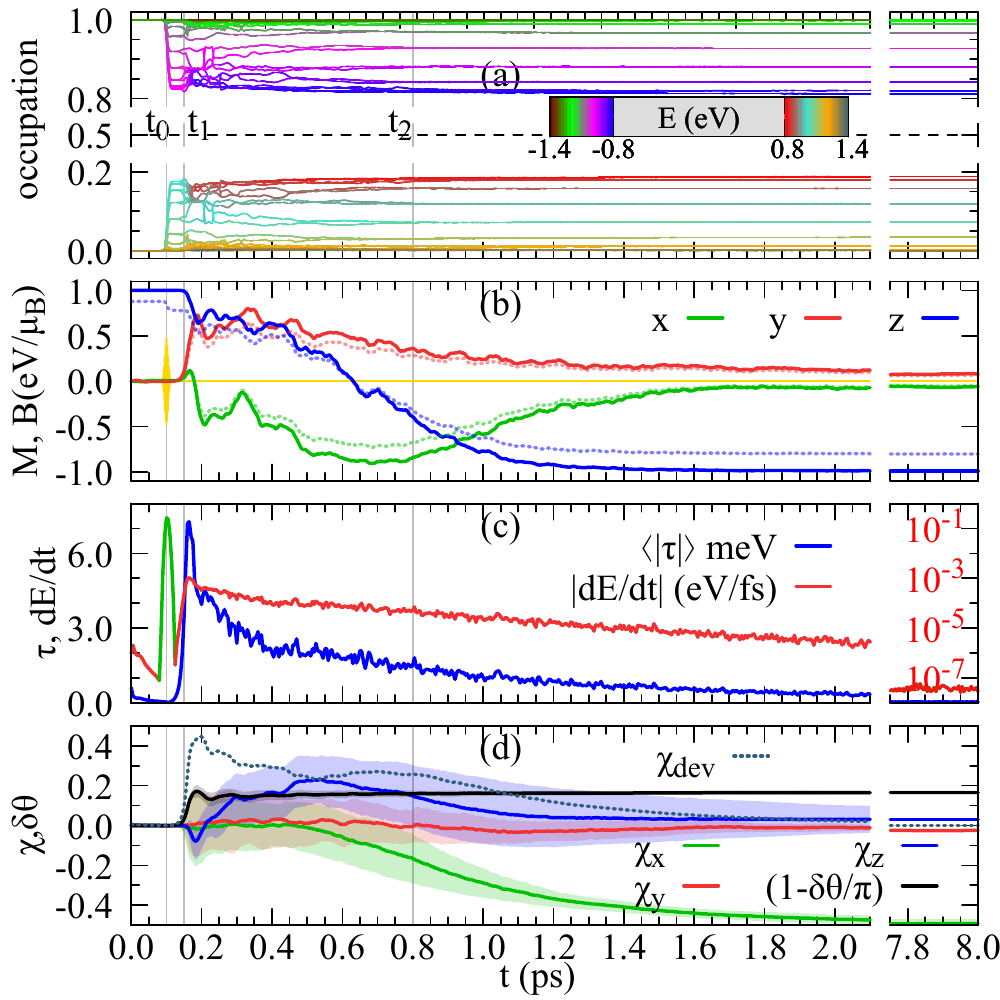}
\caption{Evolution of electron and spin subsystems over time. The laser hits at $t=t_0$, the magnetisation dynamics starts at $t=t_1$ and the slow relaxation dynamics starts at $t=t_2$.
(a) Occupation of electronic states over time at half filling. The colors correspond to the eigenvalues (shown in inset). Black dashed line shows the average occupation which remains at 0.5.  (b) Time evolution of magnetisation $\hat{\mathbf{M}}$ (solid) and effective field $\mathbf{B}$ (transparent, dotted) for 17th site. Yellow line denotes the laser pulse. (c) Average magnitude of the torque ($\langle |\mathbf{\tau}| \rangle = 1/N \sum_i^N |\mu_B \hat{\mathbf{M}}_i \times \mathbf{B}_i|$ in blue, scale on the left) and rate of change of energy  ($dE/dt$, $\log_{10}$ scale on the right). Green and red color correspond absorption and dissipation of energy. (d) Evolution of average chirality components and relative angle over time. The shaded regions show the rms deviation.}
\label{fig:occm}
\end{figure}

The interaction of the pulse with the spin system is modelled by a Peierls substitution in the kinetic hopping, assuming that there is no absorption of angular momentum, which is found to have negligible impact on the magnetisation dynamics \cite{Chen2019}.  This results in the following time-dependent Hamiltonian
\begin{eqnarray}
H(t) &=& -J \sum_{i;\mu,\nu} c_{i,\mu}^\dagger (\hat{\mathbf{M}}_i(t) \cdot \boldsymbol{\sigma} )_{\mu\nu} c_{i,\nu}  \nonumber \\
&& - h \sum_{ \langle i,j \rangle;\mu } c_{i,\mu}^\dagger c_{j,\mu} e^{i\frac{e}{\hbar} \mathbf{A}(t) \cdot \mathbf{d}_{ij}}
\label{ht} 
\end{eqnarray}
where $e$ is the electronic charge and $\mathbf{d}_{ij}$ is the vector connecting sites $i$ to $j$. The corresponding time evolution of the quantum states is evaluated in small time steps ($\delta t=0.001$\,fs). As magnetisation dynamics is a slower process as compared to the evolution of quantum states, the magnetic moments practically remain constant during each step of electronic evolution. In that case the time evolution of the $n$'th instantaneous eigenstate $|\phi_n(t)\rangle$ of $H(t)$ reads $|\phi_n(t+\delta t) \rangle = e^{-iH(t)\delta t/ \hbar} |\phi_n(t) \rangle$. Any initial state can be expressed as a linear combination of these basis states and  at any instance $t$ the time evolution of any arbitrary state $|\Psi_l\rangle$ can be expressed as $|\Psi_l(t+\delta t)\rangle = \sum_n c_{n}^{l}(t) e^{-i H(t)\delta t/ \hbar} |\phi_n(t) \rangle$, where $c_{n}^{l}(t)=\langle \phi_n(t) |\Psi_l(t)\rangle$ is the instantaneous overlap integral subjected to the initial condition $c_n^l(0) = \delta_{nl}$. This automatically ensures half filling as long as the system does not deviate too much from an antiferromagnetic ground state and positive and negative energy states remains separated by the exchange gap (Fig.\,\ref{fig:occm})a, \cite{Suppl}.

The quantum evolution of the states is coupled to the magnetisation dynamics of the  chain as governed by a set of Landau-Lifshitz-Gilbert equations
\begin{equation}
\frac{d\hat{\mathbf{M}}_i}{dt} = -\gamma' [\hat{\mathbf{M}}_i \times \mathbf{B}_i(t)] - \lambda [\hat{\mathbf{M}}_i \times (\hat{\mathbf{M}}_i \times \mathbf{B}_i(t))], \label{llg}
\end{equation}
where $\gamma' = \frac{\gamma}{1+\alpha^2} \frac{1}{\mu_i}$ and $\lambda = \frac{\gamma \alpha}{1+\alpha^2} \frac{1}{\mu_i}$ with $\gamma = \frac{g_e \mu_B}{\hbar}$, $g_e=2$ is the gyromagnetic ratio, $\mu_B$ is Bohr magneton and $\mu_i$ is the magnetic moment at site $i$ which we choose to be 1\,$\mu_B$ on each site.  $\alpha$ is the dimensionless damping parameter which we keep constant at a value of 0.2. One should note that the exact mechanism of energy dissipation in such driven system can be very complicated \cite{Aoki2014, Fotso2020, Ikeda2020}. For simplicity, we do not include the effects of relaxation due to electron-electron, electron-phonon or electron-magnon interaction explicitly, noting that they  can be effectively incorporated via a suitable modification of the damping parameter \cite{Karakurt2007, Thonig2014}.
The  exact  treatment  of  the  electron  excitation  during  the  action  of  the  laser  pulse  is important to  compute  the  nonequilibrium  spin  polarisations that  drive  the system  into  the  noncollinear  state. We assume  that  once  the  nonequilibrium spin polarisations are present and the laser-pulse has been switched off, the exact energies of the conduction electrons are less important for the magnetisation dynamics, and therefore we assume that we may ignore the fast fs energetic relaxation of the electrons. In particular, we assume that the laser excites the conduction electrons to higher energies and to different spin expectation values and that the nonequilibrium spin expectation values persist to the ps time scale while only the energies of the excited electrons relax on the fs time scale.

In Eq.~(\ref{llg}) the torque on the magnetisation, $\mathbf{\tau}_i=\mu_B \hat{\mathbf{M}}_i\times\mathbf{B}_i$, is exerted by the time dependent effective magnetic field  $\mathbf{B}_i(t)$ acting on the moment at site $i$, defined as 
\begin{eqnarray}
\mathbf{B}_i(t) = \frac{1}{\mu_B}\sum_l \langle \Psi_l(t) | - \boldsymbol{\nabla}_{\bm{M}_i} H(t) | \Psi_l(t) \rangle.
\label{bt}
\end{eqnarray}
By solving  Eqs.~(\ref{llg}) and (\ref{bt}) simultaneously we are thereby able to study the effect of the laser pulse on the magnetisation dynamics as mediated by excited electronic states. In the past, similar approach has been adopted to study the quantum evolution of magnetic system \cite{Koshibae2009, *Ono2017, Ono2019} and current driven magnetisation dynamics \cite{Petrovic2018,*Suresh2020}, however, a steady chirality formation was not observed. Ishihara and Ono \cite{Ono2019} observe a transient skyrmion-like configuration during a periodically-driven laser-assisted transition from a ferromagnetic to an antiferromagnetic state, stable on the scale of of hundred femtoseconds. Contrary to that, as discussed below, we observe formation of steady chiral states which can survive up to several picoseconds and can be generated with a finite pulse.
 
To describe the laser-excited spin and electron dynamics of the system we start with an initial AFM chain in $zx$ plane (Fig.~\ref{fig:spiral}a). The effect of finite fluctuations is incorporated by adding  small random polar ($\theta_{ran}=0.01$\,rad) and  azimuthal angle ($\phi$ randomly chosen between 0 and 2$\pi$) to each magnetisation direction. These small non-collinear components are imperative to initiate the dynamics (see~\cite{Suppl}). To describe the time evolution, we consider one specific random configuration (Fig.~\ref{fig:occm}), while the averaged behaviour is discussed in \cite{Suppl}.

\section{Results}

The dynamics of the system can be roughly divided into three different time regimes (vertical gray lines in Fig.~\ref{fig:occm}) marked by the incident of the laser ($t_0=0.1$\,ps), onset of magnetisation dynamics ($t_1=0.15$\,ps) and the onset of the slow relaxation ($t_2=0.8$\,ps). In the first regime ($t_0<t<t_1$) the occupation of different quantum states changes instantly (Fig.~\ref{fig:occm}a) without any change in the magnetisation (Fig.~\ref{fig:occm}b). The mixing of quantum states causes a change in the onsite $\mathbf{B}_i$ for the components parallel to the $\mathbf{M}_i$ only (Fig.~\ref{fig:occm}b). Being stimulated by the initial random non-collinear components, the torques starts building up around $t\sim t_1$. A stronger randomness causes a faster response \cite{Suppl}. In absence of any external stimulus, the system would eventually go back to its AFM ground state which is reflected in initial energy dissipation (Fig.~\ref{fig:occm}c) and reduction of effective torque (Fig.~\ref{fig:occm}c,\,\cite{Suppl}).

In the second regime ($t_1<t<t_2$) the system reorganises itself. The quantum states strongly mix with each other (Fig.~\ref{fig:occm}a) driving a significant change in the torques (Fig.~\ref{fig:occm}c) and the orientation of magnetic moments (Fig.~\ref{fig:occm}b).
The overall duration of this process is of the order of that typical for laser-induced demagnetisation ~\cite{Zhang2020}, which ensures that our model correctly captures the governing interactions between the laser field and magnetic moments.
After that the system enters into the third regime ($t>t_2$) where the changes becomes much slower. At $t\sim 2$\,ps, which is the same order as the reorientation time for ultrafast demagnetisation as well~\cite{Zhang2020}, the system has achieved its steady chiral state which survives on the time scale of several picoseconds~ (Figs.~\ref{fig:spiral}b) \cite{SupplMov}. Note that at 8\,ps, the system is dissipating energy at approximately $10^{-7}$\,eV/fs. Considering the energy gain from the laser to be of the order of eV, the system can dissipate the additional energy on the nanosecond scale and come back to its initial ground state. It is worth mentioning that the Gilbert damping is not the only source of energy dissipation. The interaction between different quantum states can further modify the dissipation channel and consequently the dissipation rate is not simply proportional to the damping parameter only \cite{Suppl}.

\subsection{Chirality and relative angle}

To characterise the evolution qualitatively, we define the average vector chirality $\boldsymbol{\chi} = \frac{1}{N-1}\sum_{i=1}^{N-1}  \hat{\mathbf{M}}_i \times \hat{\mathbf{M}}_{i+1} $ and angular deviation $\delta \theta = \frac{1}{N-1}\sum_{i=1}^{N-1}  \cos^{-1}  \hat{\mathbf{M}}_i \cdot \hat{\mathbf{M}}_{i+1} $,
and evaluate these quantities over time for the chosen initial state (Fig.~\ref{fig:occm}d). Since we are starting from an initial anti-ferromagnetic state, in the following we present the normalised angular deviation $1-\delta\theta/\pi$ which is zero for a perfect antiferromagnet. The \textit{perfection} of the state is quantified by the spiral deviation factor $\chi_{dev} = \left| \sin(\delta \theta)-|\chi| \right|$, which is zero for a perfect spiral. One can readily see that the chirality components and  angular deviation follow the same profile as the torque. For $t>t_2$ they converge to their asymptotic value with a decreasing rms deviation indicating an onset of the uniform spiral. This is further confirmed by the vanishing magnitude of $\chi_{dev}$. Since $\delta \theta$ relies on the relative orientation of the adjacent sites only, it saturates much faster. It takes much longer, on the other hand, to establish and saturate an averaged chirality.  The mechanism can be seen as a result of out-of-equilibrium chiral interactions such as Rashba-Bychkov interaction leading to asymmetric exchange interaction like DMI \cite{Kundu2015, Ado2018}. The saturation time of these entities depends on different system parameters~(\cite{Suppl}), with their qualitative behaviour remaining similar.

\subsection{Initial random orientation and edge effect}

\begin{figure}[h!]
\centering
\includegraphics[width=0.48\textwidth]{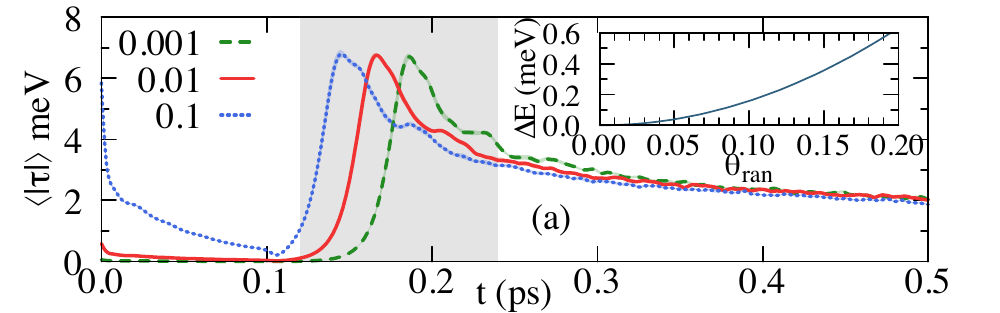}
\includegraphics[width=0.48\textwidth]{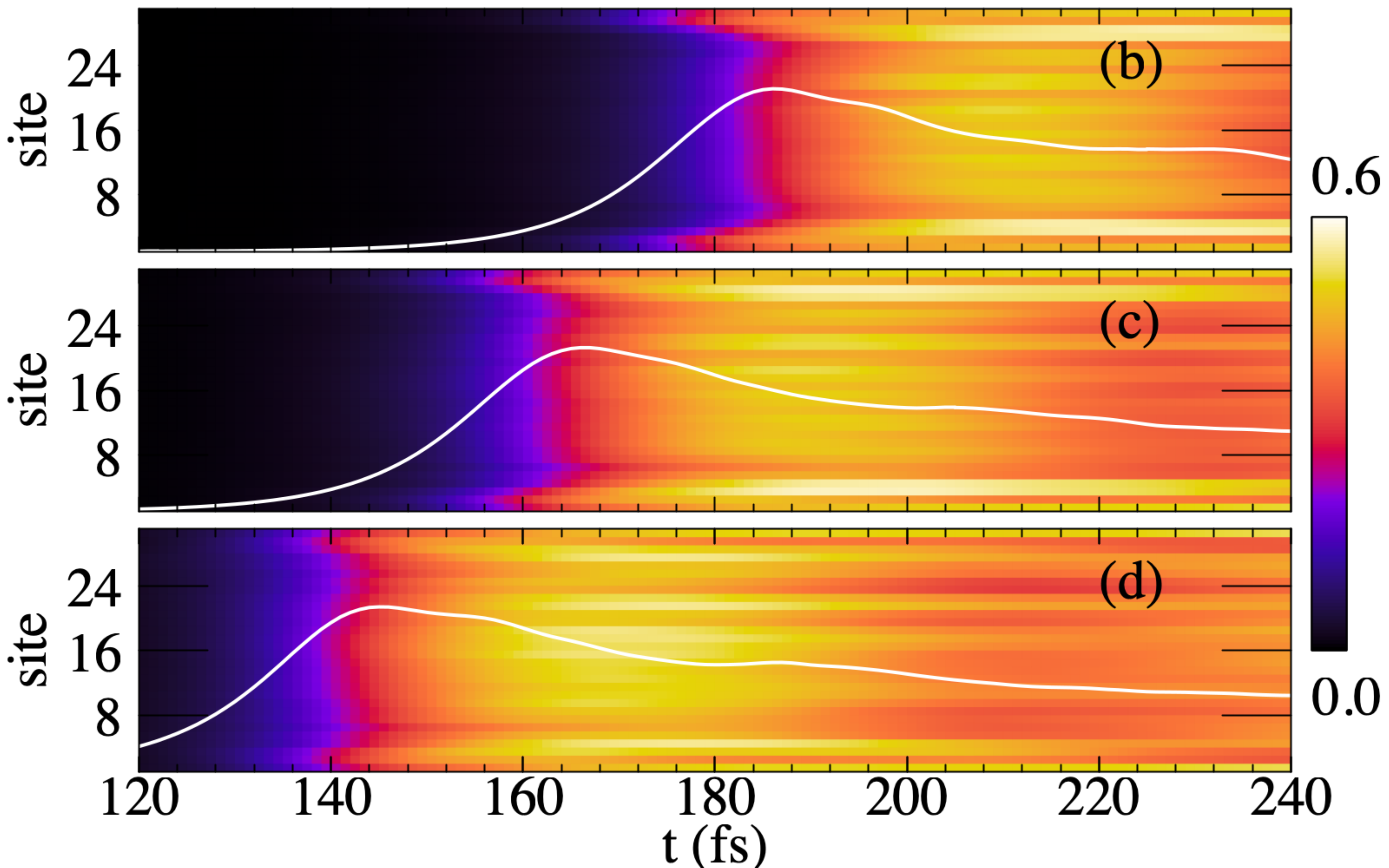}
\caption{Impact of randomness and edge efect. (a) Variation of effective torque as a function of initial randomness. Inset shows the average additional energy gain due to the randomness. Gray region denotes time span for which average chirality is calculated in (b,c,d). Evolution of magnitude of local chirality $|\chi_i|$ for an initial randomness (b) 0.001, (c) 0.01 and (d) 0.1. The plots show the distribution averaged over 64 different initial configurations. The white lines show the variation of $\langle | \tau | \rangle$ shown in (a).}
\label{fig:ran_edge}
\end{figure}

There are two driving factors which initiate the spiral formation. First is the initial randomness of magnetic orientation, which provides the initial non-chiral spin mixing interaction and thus ignites the magnetisation dynamics. In absence of any intrinsic spin-mixing interaction such as spin-orbit coupling, the magnetisation dynamics cannot be ignited without a small non-collinear magnetic component. A stronger randomness promotes more mixing which results in a faster build-up of effective torque (Fig.\,\ref{fig:ran_edge}a) when compared to the case with smaller randomness. This behaviour is consistent with the time-dependent density functional theory based finding of ultrafast demagnetisation \cite{Chen2019} where a stronger randomness results in a faster demagnetisation. As described in introduction, the randomness is introduced as a small random polar angle $\theta_{ran}$ which pushes the system away from its actual ground state. This is reflected in the total energy of the system (Fig.\,\ref{fig:ran_edge}a inset). Initially the system tries to come back to its true collinear ground state which is noticeable in the initial decrease in the torque (Fig.\,\ref{fig:ran_edge}a). After being hit by the laser pulse at 0.1ps, the consequent mixing of quantum states causes a build up in torques (Fig.\,\ref{fig:ran_edge}a) and initiates the magnetisation dynamics which is reflected in the rising magnitude of the average chirality (Fig.\,\ref{fig:ran_edge}b,c,d). In pace with the torques the build up of chirality is also intensified by  stronger randomness. Note that while the evolution of individual components of chirality can be different for different initial configurations \cite{Suppl}, their magnitudes show a common trend which demonstrates the impact of the edges as the second governing factor. The edges act as a source of inversion symmetry breaking and thus can promote a dominant component of chirality as well \cite{Suppl}. The pivotal role of the edges is also reflected in the fact that the spiralisation starts at the edges $-$ an effect which is more pronounced for smaller randomness (Fig.\,\ref{fig:ran_edge}b,c,d). For larger randomness the spiralisation, on the average, can initiate from any part of the system. 

It is worth mentioning that in the present study we are not considering any structural defects of the underlying lattice. Such defects can give rise to new features in the optical response of a system and can play an important role in the outcome of any photo-induced experiment. In one dimension, the presence of defects can be of crucial importance for establishment of chiral coherence and dynamics of chirality nucleation along the chain, via the effect similar to that of the edges, where the changes in local environment can promote a faster build up of specific chirality. In higher dimensions, we believe that various types of defects  can be exploited in a similar fashion to nucleate domains of specific chirality with optical pulses, in a way similar to that of defect- or edge-assisted skyrmion nucleation with currents or pulses of magnetic field.

\subsection{Chiral spin mixing interaction}

\begin{figure}[t!]
\centering
\includegraphics[width=0.49\textwidth]{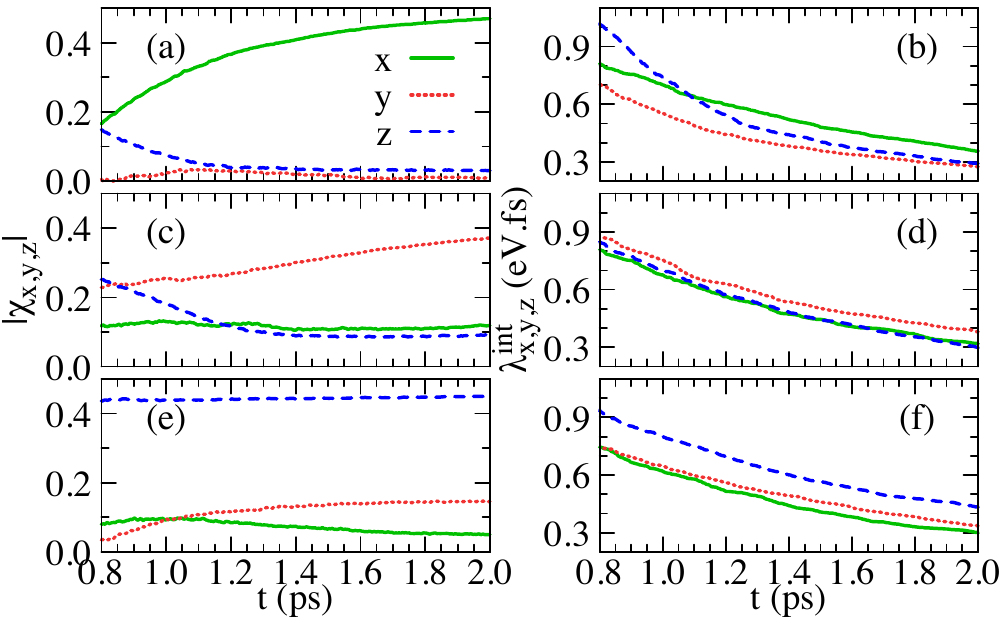}
\caption{Correlation between the components of $\mathbf{\chi}$ ($a,c,e$) and integrated CSMI strength $\mathbf{\lambda}^{int}$ ($b,d,f$). ($a,b$) correspond the configuration considered in Figs.~\ref{fig:spiral}-\ref{fig:occm}. ($c$,$d$) and ($e$,$f$) show the correlation for two other initial configurations . The green, blue and red lines in ($a$,$c$,$e$) show the magnitude of $\chi_{x,y,z}$ and in ($b,d,f$) they show the values of $\lambda^{int}_{x,y,z}$.}
\label{fig:chi-lam}
\end{figure}

Traditionally, the emergence of chirality is attributed to some sort of chiral spin mixing interaction (CSMI) (e.g. of Rashba-Bychkov type) leading to an antisymmetric exchange interaction (e.g. of DMI type). While our model itself does not possess any kind of CSMI intrinsically, the mixing of quantum states can give rise to an emergent CSMI. Similar mechanism is widely adopted in simulating SOC effect with ultra cold atoms \cite{Vaishnav2008, Liu2009}. In the latter case very few quantum states  are manipulated with a monochromatic laser which makes it possible to manipulate the exact form of the SOC \cite{Wu2016}. In contrast to that, in our case many more quantum states are involved and they are stimulated by a short pulse which excites many transitions as observed in the change of occupation of different quantum states (Fig.~\ref{fig:occm}a). To validate the existence of any emergent CSMI we consider the most generic form of staggered (since we are considering antiferromagnetic configuration) chiral spin-mixing Hamiltonian, given by $H_\lambda^r=\sum_{\langle jk \rangle} (-1)^j (i c^\dagger_j \sigma_r c_k - i c^\dagger_k \sigma_r c_j )$ for $r=x,y,z$. Note that for our linear chain $H_\lambda^y$ defines a staggered Rashba-Bychkov interaction which generates the $y$ component of DMI. For our finite chain we define the instantaneous chiral spin-mixing  strength as $\lambda_{r}(t) = \sum_l \langle \Psi_l(t)| H_{\lambda}^r | \Psi_l(t) \rangle/N$ and evaluate it over time (Fig.~\ref{fig:chi-lam}a$_1$). 

To characterise the effective contribution of these interactions we define the integrated chiral strength as $\lambda_r^{int}(t)=\int^{t_s}_t|\lambda_r(t')|dt'$, where $t_s$ corresponds to the time when the system reaches its steady state (8\,ps in our case). To clarify our point, we consider three different initial configurations where the $x$, $y$ or $z$ component of $\bm{\chi}$ dominates in the steady state (green, blue and red lines in Fig.~\ref{fig:chi-lam}) and compare corresponding $\lambda_r^{int}(t)$.  For brevity, in Fig.~\ref{fig:chi-lam} we show the evolution  of the latter quantity within 0.8$-$2\,ps when the initial turbulence is over and the system starts moving towards its final configuration. One can readily see that the components of both $\bm{\chi}$ and $\lambda^{int}$ shows same trend. For example both in Fig.~\ref{fig:chi-lam}$a_1$ and Fig.~\ref{fig:chi-lam}$b_1$ the $x$ components gradually becomes dominant while the other two components remains close. In Fig.~\ref{fig:chi-lam}$a_2,b_2$ the $y$ component is larger while Fig.~\ref{fig:chi-lam}$a_3,b_3$ is marked by a dominant $z$ component. The correspondence between the components of effective $\lambda^{int}$ and that of the chirality therefore establishes the emergent CSMI as the source of the induced chirality. The emergent CSMI can be thus considered as the source of an out-of-equilibrium type of anti-symmetric exchange interaction, such as the Dzyaloshinskii-Moriya interaction, leading to a spiral in accord with  equilibrium scenario~\cite{Koralek2009, Yang2018}. Due to the random nature of the initial states, the system can promote any random combination of emergent interactions leading to any random end chirality. A preference can arise from the initial magnetic polarisation which can cause a particular combination of quantum states to appear in abundance~\cite{Suppl}. Further preference can be exerted by initially introducing a small spin dependent hopping of  Dresselhaus or Rashba-Bychkov type, which can maximise a particular flavor of mixing resulting in a specific end chirality~\cite{Suppl}. Tuning the initial polarisation and spin dependent hoping thus provides a controllable way to manipulate the end configuration.

\subsection{Dependence on laser pulse intensity}

\begin{figure}[t!]
\centering
\includegraphics[width=0.48\textwidth]{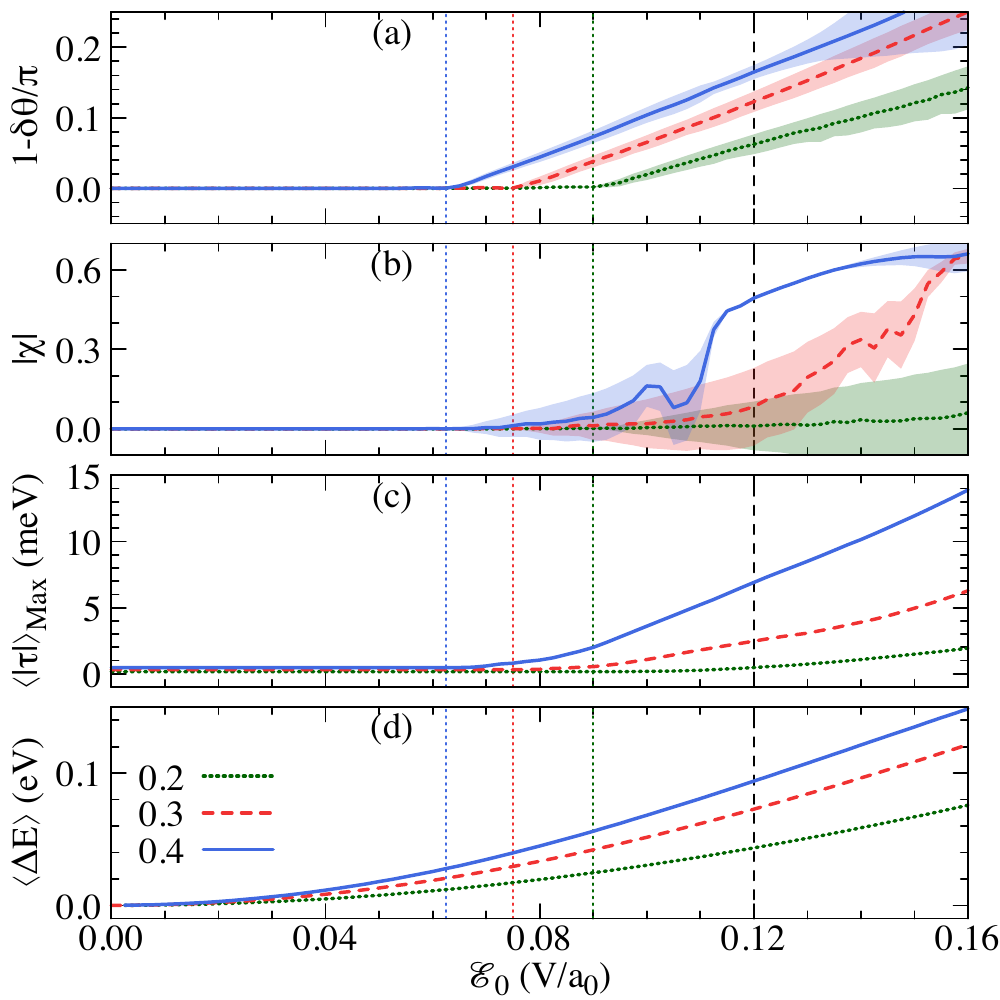}
\caption{Variation of different observables with respect to applied field strength for different hopping parameters (green 0.2eV, red 0.3eV and blue 0.4eV). Variation of (a) relative angle ($\delta \theta$), with the rms deviation given by shaded region; (b) chirality ($|\chi|$), with the magnitude of $\chi_{dev}$ given by the width of the shaded region; (c) maximum average torque ($\langle |\tau| \rangle_{Max}$), and (d) average pumped energy ($\langle \Delta E \rangle$)  with respect to the amplitude of applied pulse ($\mathcal{E}_0$). Vertical green, red and blue lines show the onset of $\delta \theta$ for corresponding hopping parameters. The values shown here are taken at 8\,ps after averaging over 16 different configurations.  Vertical black dashed line correspond electric field used in Fig.~\ref{fig:occm}.}
\label{fig:chi-th}
\end{figure}
 
A natural question arises whether it is possible to manipulate  the degree of chirality in the system. In absence of any bias  we characterise a spiral with $|\chi|$ and $\delta \theta$ and the deviation from a perfect spiral is given by the quantity $\chi_{dev}$. From Fig.~\ref{fig:chi-th}a,b one can see that both $\delta \theta$ and $|\chi|$ pick up a finite value after a threshold field strength, which in turn correlates with the onset of the torque (Fig.~\ref{fig:chi-th}c), and decreases with the increase in hopping parameter which facilitates strong overlap between sites. Note that even before the threshold the system absorbs energy from the pulse (Fig.~\ref{fig:chi-th}d), which is invested into electronic transitions across the gap and changes quadratically with the pulse strength ($\mathcal{E}_0$). After the threshold the  pumped energy is also partly utilised for forming chirality, which results in a deviation from the quadratic nature of the dispersion (Fig.~\ref{fig:chi-th}d).  Close to the threshold, large  values of $\chi_{dev}$, Fig.~\ref{fig:chi-th}b,  suggest that although the adjacent local magnetic moments are having a uniform angular deviation, the overall structure lacks global spiral order. Away from the threshold, the chain forms a steady spiral as marked by the vanishing $\chi_{dev}$. For much stronger pulse intensities  the system deviates from the steady spiral state again. This is due to the fact that a pulse with higher amplitude pumps more energy and the system requires more time to attain its smooth spiral configuration for fixed damping constant. Similar behaviour is observed with a increase of duration of the pulse or the frequency of the laser \cite{Suppl}.

\subsection{Ambient temperature}

\begin{figure}[t!]
\centering
\includegraphics[width=0.48\textwidth]{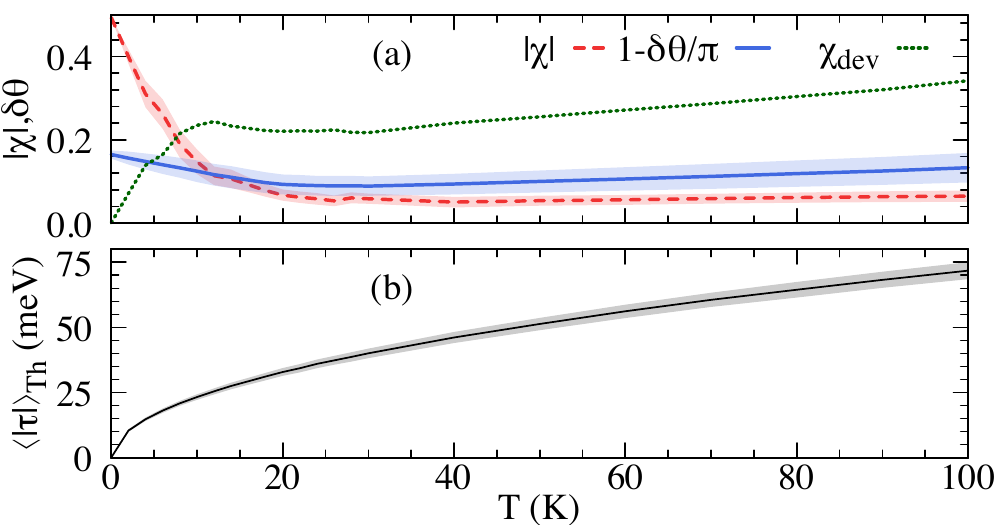}
\caption{Effect of ambient temperature. Variation of (a) chirality ($\chi$ in red), relative angle ($\delta \theta$ in blue), $\chi_{dev}$ (green) and (b) average thermal torque, $\langle |\tau| \rangle_{Th}$ (rms deviation as shaded region), with temperature, for hopping parameter of 0.4 eV. The values are obtained at 8 ps from averaging over 64 configurations over a time period of 50 fs.}
\label{fig:Th}
\end{figure}

We finally address the effect of ambient temperature on the stability of the laser-driven spiral states, which in our calculations is introduced as an additional time-dependent onsite random field $\bm{B}^{Th}_i(t) = \sqrt{(2\alpha k_B T \mu_i /\gamma \delta t)} \bm{\eta}_i(t)$, where each component of $\bm{\eta}(t)$ follows a normal distribution over the sites at each time \cite{Muller2019}. Considering the small system size, this approach may lead to an overestimation of thermal effects. Due to the large (2\,eV) gap, the ambient temperature range under consideration would not cause any change in electronic occupation. With increasing $T$, chirality decreases rapidly away from its ``ideal" value, whereas the angular deviation exhibits a very robust behavior~(Fig.~\ref{fig:Th}a). However, chirality does not vanish entirely but rather saturates at a residual value, although the magnitude of $\chi_{dev}$ clearly indicates that the resulting state is far from being a perfect spiral. It is important to realize that the torque due to the thermal fluctuations, Fig.~\ref{fig:Th}b, does not depend on model parameters, whereas the effective field and corresponding torque do (Fig.~\ref{fig:chi-th}c). Therefore it might be possible to reduce the thermal impact by tuning the system parameters.

\section{Discussion}
In our work we introduce an overlooked paradigm to imprint chirality in magnets with a finite laser pulse. Our results clearly show that the laser-driven magnetisation dynamics is preceded by the electronic excitation with a lag time depending on the initial magnetic configuration. By establishing a bridge between the quantum evolution of electronic states and classical spin dynamics, we present a  comprehensive picture of the laser-mediated formation of large-scale chiral states. It is important to realize, that the uncovered mechanism is distinctly different from that which is commonly used to interpret 
laser excited dynamics in terms of transfer of effective temperature between electronic to spin degrees of freedom which is also known as two temperature or three temperature model (considering the phonon modes) \citep{Atxitia2010, Chimata2012}. Although these phenomenological models works nicely for ultrafast demagnetisation, they fail to capture the physics of ultrafast generation of chirality. 

\begin{figure}[h!]
\centering
\includegraphics[width=0.35\textwidth]{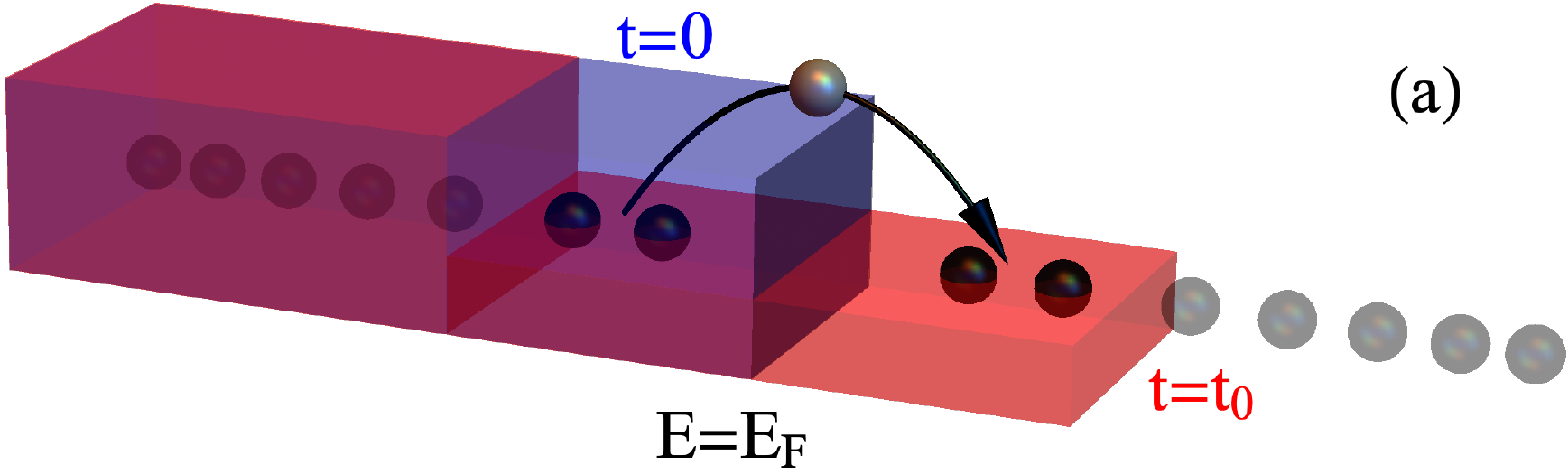}
\includegraphics[width=0.48\textwidth]{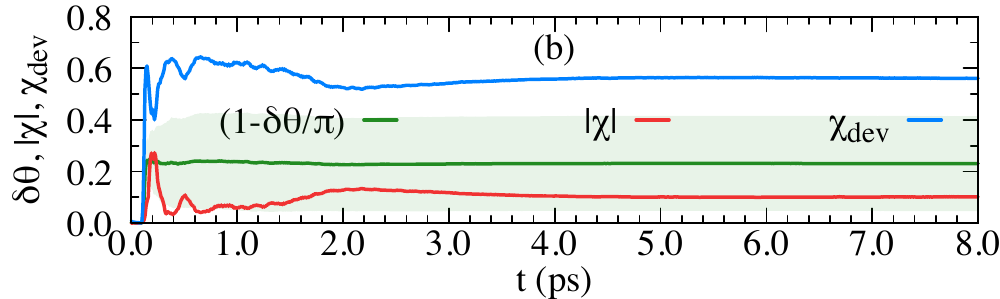}
\includegraphics[width=0.48\textwidth]{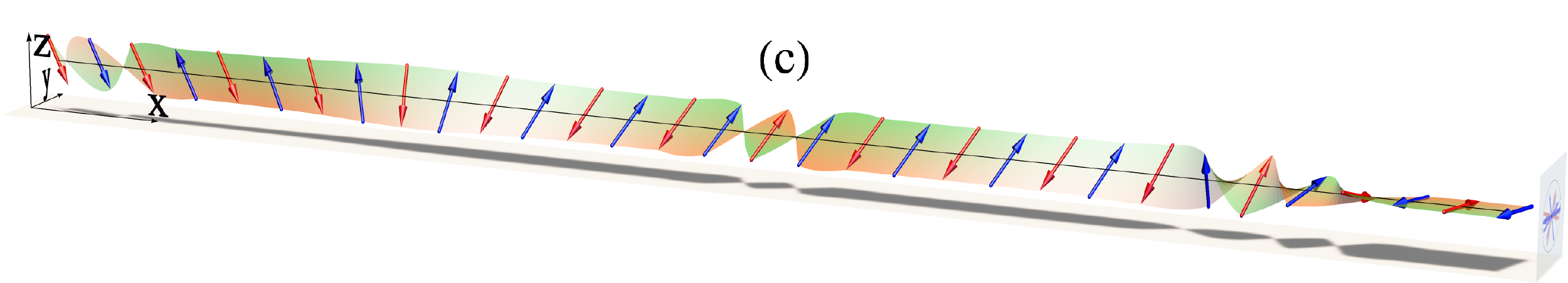}
\caption{ Evolution of thermally initiated magnetisation dynamics. (a) Schematic of simulating thermal occupation. Blue and red regions show the occupation distribution at $t=0$ and $t=t_0$ which is introduced to simulate the thermal impact of the laser. (b) Relative angle ($\delta \theta$), chirality ($|\chi|$) and chiral deviation ($\chi_{dev}$) for for a specific configuration. Shaded green region shows the rms deviation of $\delta \theta$. The excitation takes place at 0.1ps. (c) End configuration of the evolution at 8\,ps.}
\label{fig:chi_th}
\end{figure}

To show this, we artificially simulate the effect of heating of the electronic sub-system. Due to the presence of a large gap and the fact that the laser pulse only changes the occupation of a small range of eigenstates within a specific energy window, it is not possible to construct a thermal distribution to capture this phenomena. Therefore we replace the laser  with a change of occupation by hand at time $t_0$ (Fig.~\ref{fig:chi_th}a) with a random phase factor to simulate the effect of thermal transition and let the system evolve (Fig.~\ref{fig:chi_th}b). Notably, we observe that with thermal excitation the system forms multiple domains rather than a smooth spiral (Fig.~\ref{fig:chi_th}c). In accord to existing knowledge, such formation can change the effective magnetic order which can be observed in experiments on ultrafast demagnetisation. However the large rms deviation of $\delta \theta$ and large values of $\chi_{dev}$ (Fig.~\ref{fig:chi_th}b) compared to laser-excited dynamics shows that the thermal excitation does not guide the system to a spiral formation. This has been further clarified by showing the end configuration of the evolution (Fig.~\ref{fig:chi_th}c, \cite{Suppl}).

The approach used in our work successfully captures the salient features of of the intertwined electronic and magnetic dynamics which is crucial in  optical manipulation of chiral magnetic structures, and paves a way for further material design aimed at optimisation of time scales and energetics involved. The proposed paradigm also provides a fruitful ground for studying diverse effects related to the interplay of optical chiral dynamics with the effects of spin relaxation, impurity scattering and phonon excitations. The ability to take into account a large number of atoms as well as related spin and electronic dynamical effects can become indispensable for addressing the influence of chiral dynamics on demagnetisation processes~\cite{Chen2019}, or optically assisted creation of exotic chiral particles such as hopfions~\cite{Rybakov2019}.   
Such versatility can be instrumental in exploring new possibilities of light-matter interaction in magnetic materials.

\section{Acknowledgements}
We thank Nikolai Kiselev, Filipe Guimar\~aes, Manuel dos Santos Dias and Samir Lounis for discussions.  
We acknowledge financial support from Leibniz Collaborative Excellence project OptiSPIN $-$ Optical Control of Nanoscale Spin Textures.
We  acknowledge  funding  under SPP 2137 ``Skyrmionics" of the DFG. We gratefully acknowledge financial support from the European Research Council (ERC) under the European Union's Horizon 2020 research and innovation program (Grant No. 856538, project "3D MAGiC”). The work was also supported also by the Deutsche Forschungsgemeinschaft (DFG, German Research Foundation) $-$ TRR 173/2 $-$ 268565370 (projects A11 and B12), TRR 288 $-$ 422213477 (projects B06 and A09).  We  also gratefully acknowledge the J\"ulich Supercomputing Centre and RWTH Aachen University for providing computational resources under project Nos. jiff40 and jpgi11.

\section{Data availability}
The data that support the findings of this study are available from the corresponding author upon reasonable request.

\section{Author contributions}
Y.M., O.G. and S.B. initiated the project. S.G., F.F. and Y.M. formulated the problem and worked out the methodology. S.G. implemented the method and performed the simulations. Y.M. coordinated the project. All authors discussed the results contributed to the analysis of the data and writing of the manuscript.

\section{Competing interests}
The authors declare no competing interests.

\bibliographystyle{apsrev4-1}
\bibliography{refs}

\pagebreak

\begin{center}
\textbf{\large Supplementary Materials:\\ Driving spin chirality by electron dynamics in laser-excited antiferromagnets}
\end{center}

\section{Dispersion relation for spin spiral}
\begin{figure}[h!]
\centering
\includegraphics[width=0.48\textwidth]{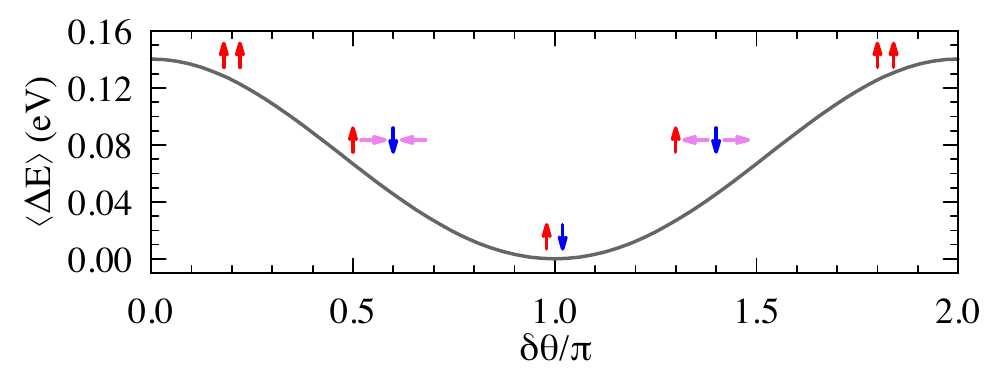}
\caption{Energy per site of the spiral as a function of spiral angle ($\delta \theta$) at half filling. The spiral is confined to the $zx$-plane and the AFM ground state is chosen to be at zero energy.}
\label{fig:dispersion}
\end{figure}

The dispersion relation of the spin spiral is useful to find out the ground state of a given model. To obtain the dispersion relation we construct a spiral in $xz$ plane for a given value of angular deviation ($\delta \theta$) and calculate the eigenvalues. The ground state energy is given by the sum of lowest half of the eigenvalues with the Fermi level remaining at zero. We choose the AFM ground state energy as zero level and calculate the dispersion for other values of $\delta \theta$ (Fig.\,\ref{fig:dispersion}). The dispersion remains the same for soirals formed in $xy$ or $yz$ plane.

\section{Evolution of magnetisation}

\begin{figure}[h!]
\centering
\includegraphics[width=0.48\textwidth]{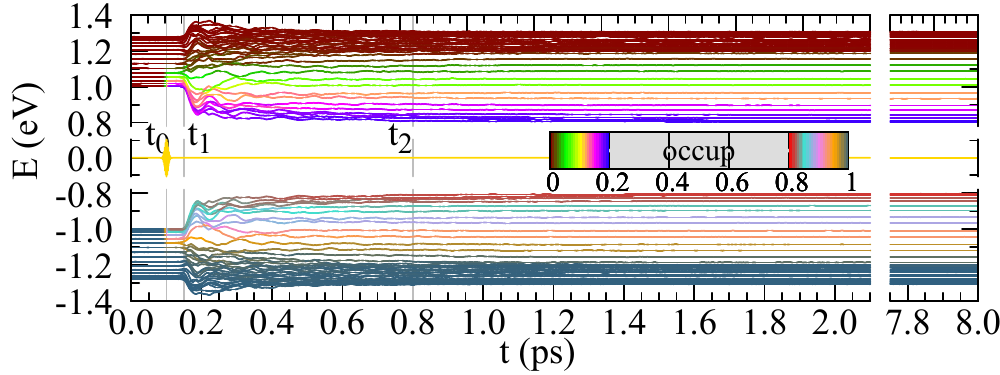}
\caption{Time evolution of instantaneous eigenvalues of the Hamiltonian with their individual occupation denoted by colors (shown in the legend).}
\label{fig:bands}
\end{figure}

\begin{figure}[h!]
\centering
\includegraphics[width=0.48\textwidth]{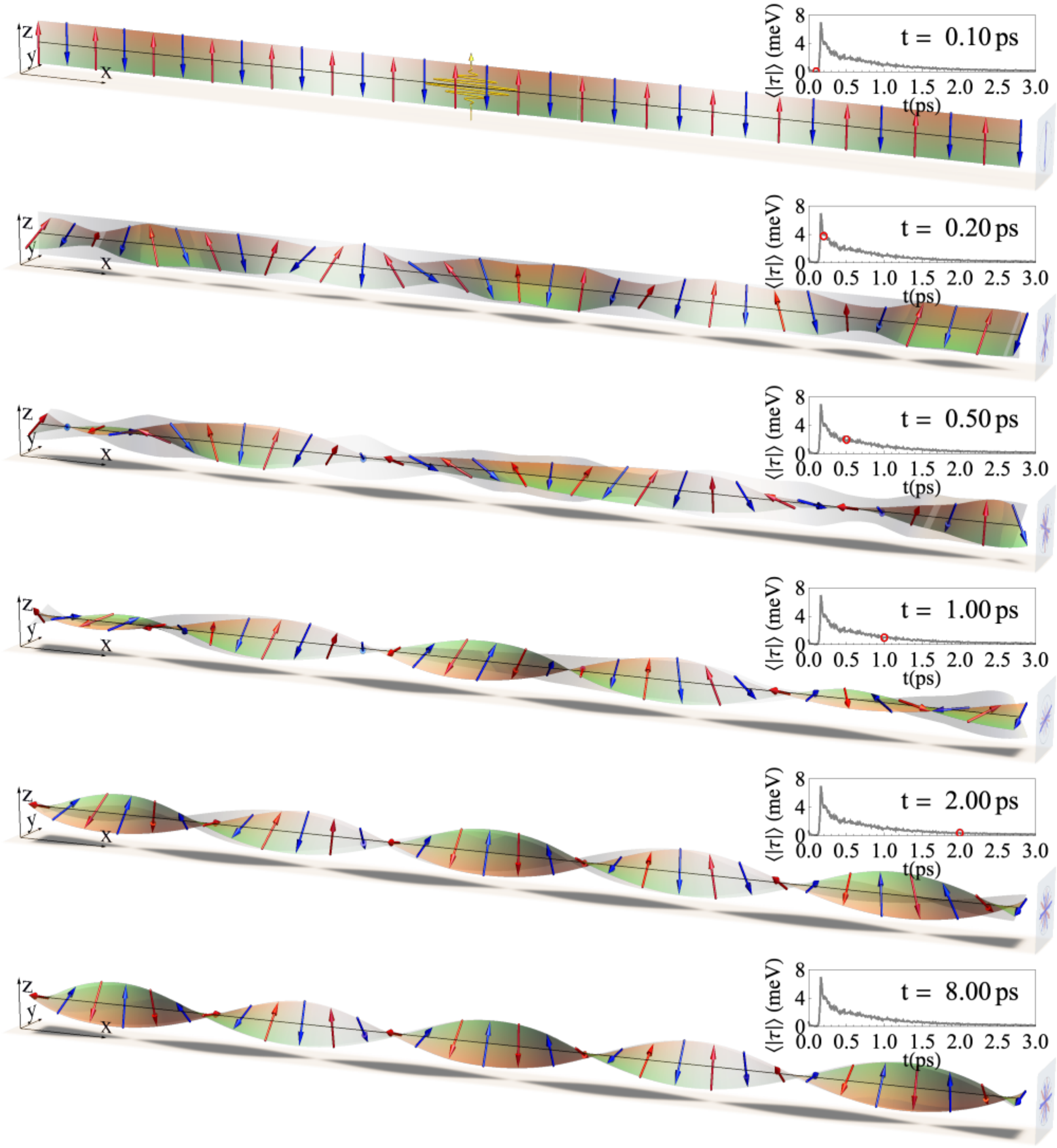}
\caption{Intermediate configurations during the formation of stable spiral. Inset shows the change in average magnitude of the torque. Transparent gray region shows the configuration of previous subfigure to distinguish the change.}
\label{fig:spiral_evol}
\end{figure}

After being hit by the laser pulse, both the electronic and magentic degrees of freedom undergo drastic changes with different characteristic timescales leading to the formation of a spiral (Fig.\,\ref{fig:spiral_evol}). While the change of electronic states is reflected in the change of the electronic occupation, the change of magnetic degrees of freedom can be traced from the evolution of instantaneous eigenvalues (Fig.\,\ref{fig:bands}). Here one can clearly see that only the states close to valence band minima / conduction band maxima are connected by the laser ($\hbar \omega$=2.02eV) and take part in the dynamics. Although the occupation changes instantaneously at $t_0$, the magnetic configuration and therefore the Hamiltonian start changing after $t_1$. The change is visible during the fast dynamics ($t<t_2$). After that the rate of change of energy becomes less than $10^{-4}$eV/fs (see Fig.\,2c in main text) and it is not visible here. Similar changes can be observed in the magnetic configuration as well (Fig.\,\ref{fig:spiral_evol}).

\clearpage

\section{Effect of initial randomness}

The initial randomness ($\theta_{ran}$), added as a small random deviation of the polar angle, plays a crucial role in initiating the magnetization dynamics. Stronger is the randomness, quicker it generates the initial non-equilibrium torque (see Fig.\,3a in main text) which eventually leads to the magnetisation dynamics. Without the initial randomness, the laser generated effective field lies parallel to the magnetic moments which does not exert any torque and consequently the magnetization dynamics does not start.

\begin{figure}[h!]
\centering
\includegraphics[width=0.48\textwidth]{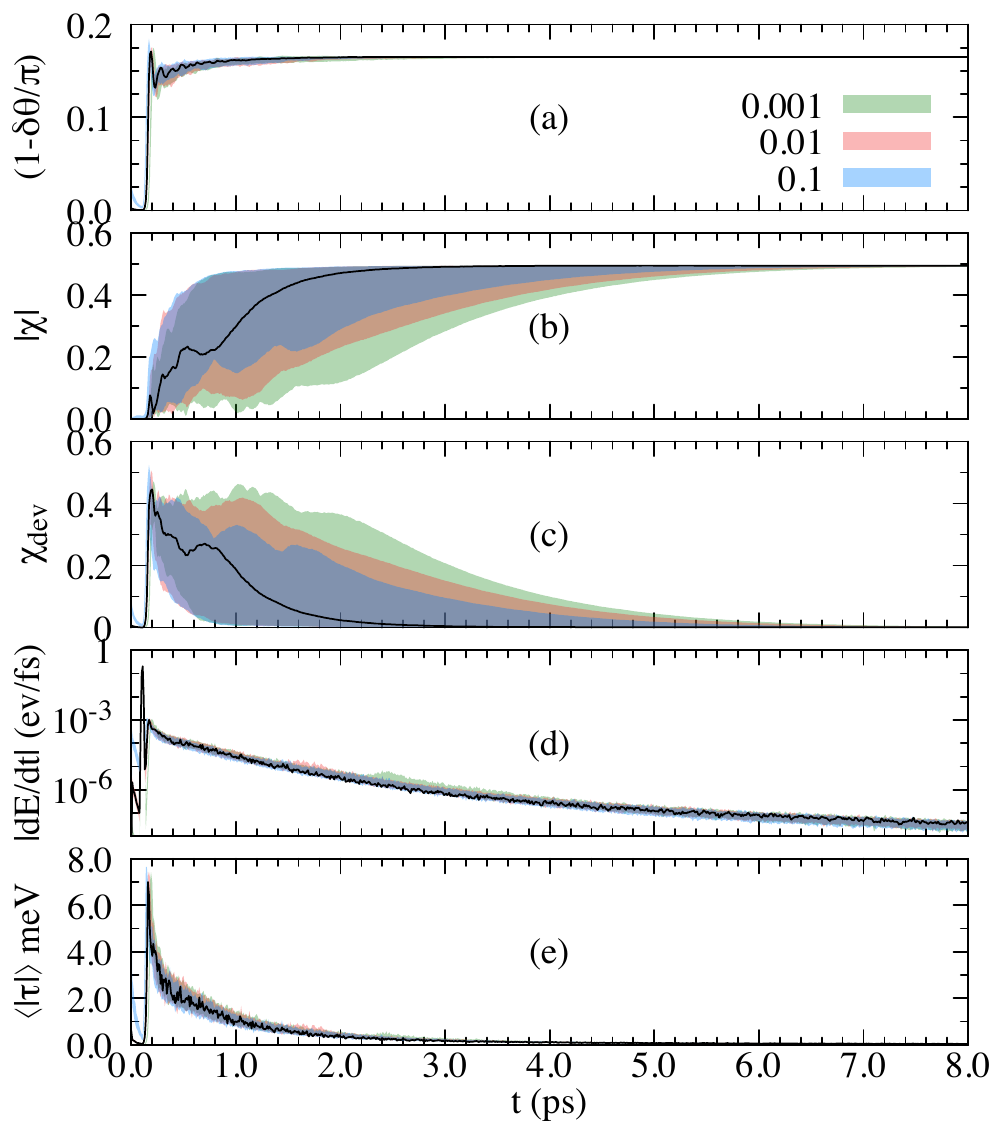}
\caption{Variation of different entities with the initial randomness. (a) Relative angle ($\delta\theta$), (b) magnitude of average chirality ($|\chi|$), (c) deviation from ideal spiral ($\chi_{dev}$),  (e) rate of change of energy in ev/fs and (f) average torque ($\langle |\tau| \rangle$) in meV with shaded regions showing the region spanned by the 64 trajectories for a given amplitude of initial randomness (given in the legend). The black line shows the trajectory of the configuration presented in the main text.}
\label{fig:avg-chi}
\end{figure}

Fig.\,\ref{fig:avg-chi} shows the impact of initial randomness on the evolution process via several physical observables. Depending on the initial random configuration the final configuration can posses any vector chirality with the same magnitude and relative angle and therefore we consider these two entities and their derivatives to characterise the effect.  With a stronger randomness all possible trajectories stay close to each other and converge faster than the case with smaller randomness. Note that on the average the relative angle ($\delta \theta$) saturates much faster than the average chirality which denotes that the relative angle is controlled by the fast electronic interaction rather than the slow magnetisation dynamics. The energetics of the evolution, on the other hand, does not have any strong influence of initial randomness, especially after the system enters the slow dynamics.

\section{Effect of damping parameter}

\begin{figure}[h]
\centering
\includegraphics[width=0.48\textwidth]{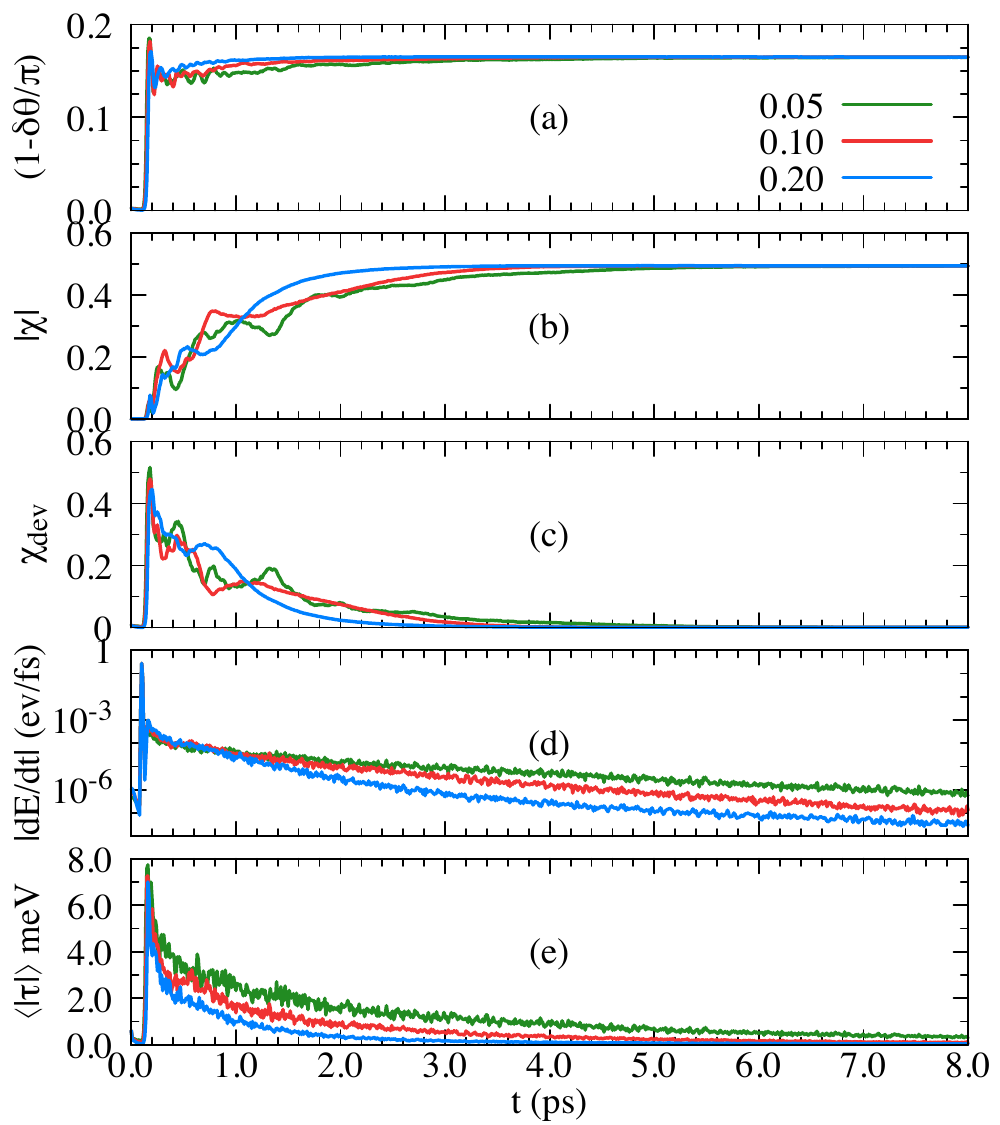}
\caption{Dependence on the damping. Evolution of (a) relative angle ($\delta\theta$), (b) magnitude of average chirality ($|\chi|$), (c) deviation from ideal spiral ($\chi_{dev}$),  (e) rate of change of energy in eV/fs and (f) average torque ($\langle |\tau| \rangle$) in meV for different damping parameters (shown in the legend). The initial configuration and the rest of parameters are the same as in the main text.}
\label{fig:damp}
\end{figure}

The LLG damping parameter controls the time frame of the spiral formation. The saturation time increases with a decrease of the damping and in absence of the damping the system never reaches a steady spiral formation. Note that damping has a smaller impact on the relative angle (Fig.\,\ref{fig:damp}a) than the chirality which implies that $\delta \theta$ is less influenced by the magnetization relaxation properties. Note that after 2\,ps, when the system enters the slow reorientation dynamics, the energy dissipation is smaller for stronger damping parameter which indicates that the LLG damping is not the governing damping mechanism here. A stronger damping drives the system faster towards its steady state and consequently, during the fast dynamics ($t\leq1$\,ps) the change in energy is proportional to the strength of the damping parameter.

%\clearpage
\section{Frequency and width of laser}

\begin{figure}[h]
\centering
\includegraphics[width=0.48\textwidth]{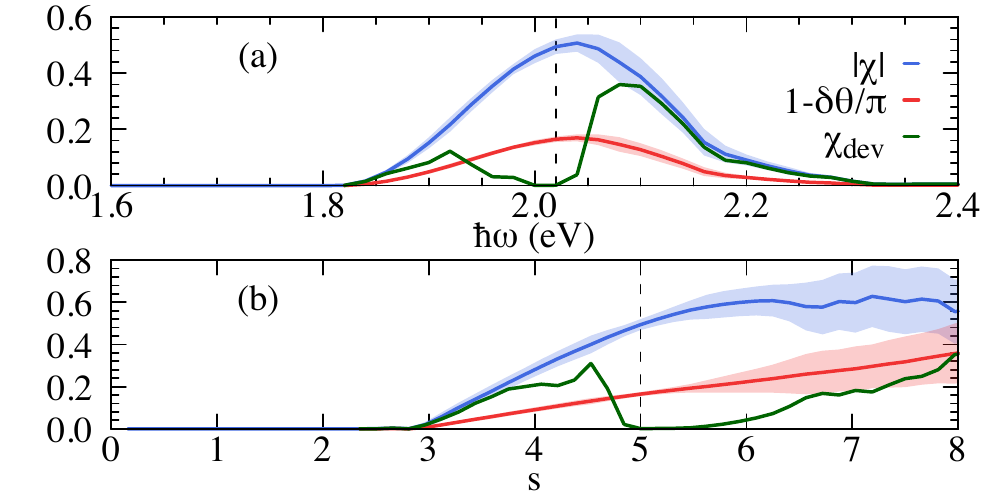}
\caption{Dependence on frequency ($\omega$) and width of the laser. The width of the laser is denoted in terms of its standard deviation $s$ where the laser pulse is represented by $\mathbf{\mathcal{E}}(t) = \mathcal{E}_0 \cos(\omega t)e^{-(t-t_0)^2/2s^2} \hat{\mathbf{x}}$. For frequency dependece the width is kept at $s=5$\,s, and for width dependence the frequency is kept at $\hbar \omega = 2.02$\,eV. The amplitude of the laser is kept at 0.12\,V/$a_0$ for both cases. The blue, red and green region shows $|\chi|$, $1-\delta \theta/\pi$ and $\chi_{dev}$ with their rms deviation denoted by the shaded region.}
\label{fig:s_omega}
\end{figure}

The dynamics is also quite sensitive to the frequency and the temporal width of the laser. The frequency determines the states which can interact via the laser while the width of the laser determines the amount of energy pumped into the system. In case of frequency, the system starts responding after the frequency crosses the initial threshold to overcome the exchange gap. Initially, the pumped energy is too small to form a smooth spiral which is reflected on the finite $\chi_{dev}$. For large frequency, the system absorbes a lot of energy which push it far from spiral state. As a result the spiral state is not achieved within 8ps and the end configuration shows a large value of $\chi_{dev}$ (Fig.~\ref{fig:s_omega}a). Similar behaviour is observed with the increase of the temporal width of the pulse (Fig.~\ref{fig:s_omega}b) as well as with the increase of amplitude of the pulse (see Fig.\,5 main text).

\section{Controling the chirality of the final configuration}

\begin{figure}[h!]
\centering
\includegraphics[width=0.45\textwidth]{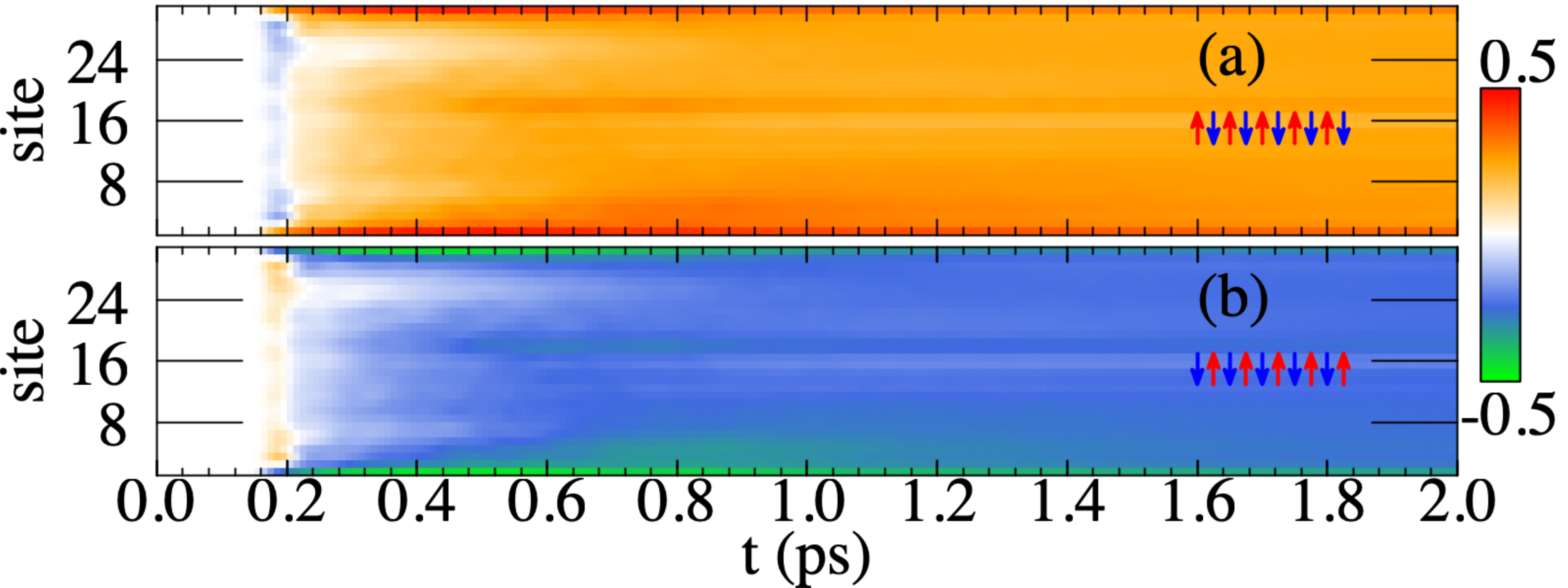}
\caption{Time evolution of $z$ component of $\chi$ for opposite magnetic configurations (Fig.~\ref{fig:chi_pol}a,b). The values are averaged over 64 different initial configurations.}
\label{fig:chi_z}
\end{figure}

\begin{figure}[ht!]
\centering
\includegraphics[width=0.45\textwidth]{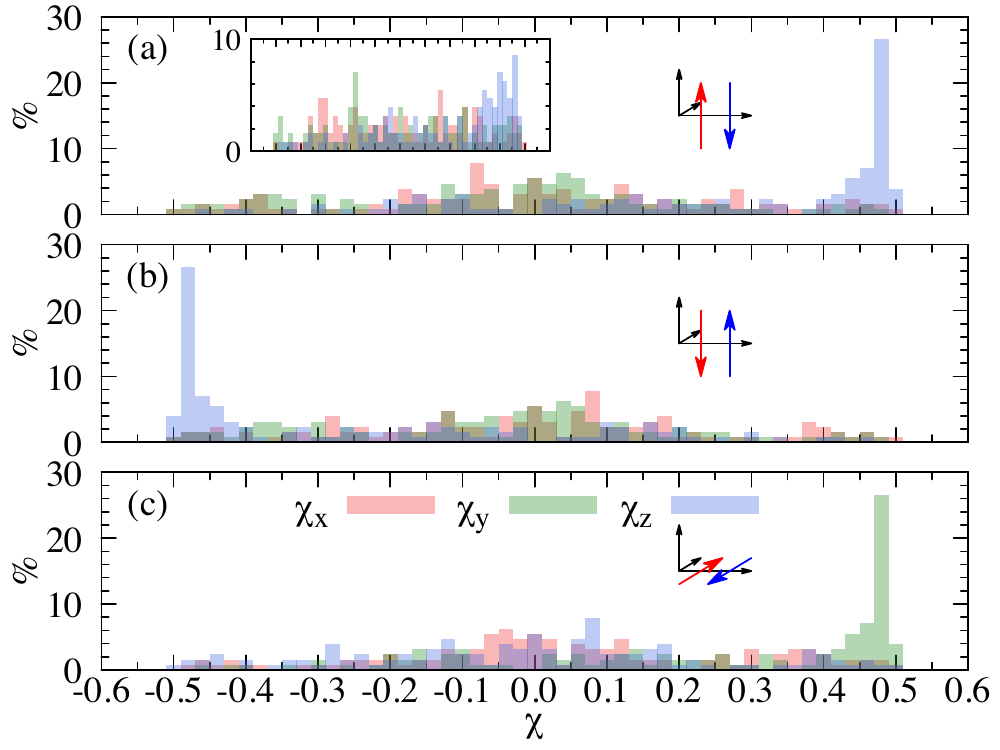}
\caption{Distribution of end chirality depending on the initial AFM configuration. The system parameters are the same as those used in the main text. (a), (b) and (c) correspond to initial AFM state with order parameter along $\hat{z}$, $-\hat{z}$ and $\hat{y}$. $x$ axis shows the value of the different components of $\chi$, and $y$ axis shows the percentage of occurrence in 128 different configurations. Inset of (a) shows the distribution with smaller damping ($\alpha=0.1$) with other parameter remaining same as that used for (a).}
\label{fig:chi_pol}
\end{figure}

\begin{figure}[ht!]
\centering
\includegraphics[width=0.45\textwidth]{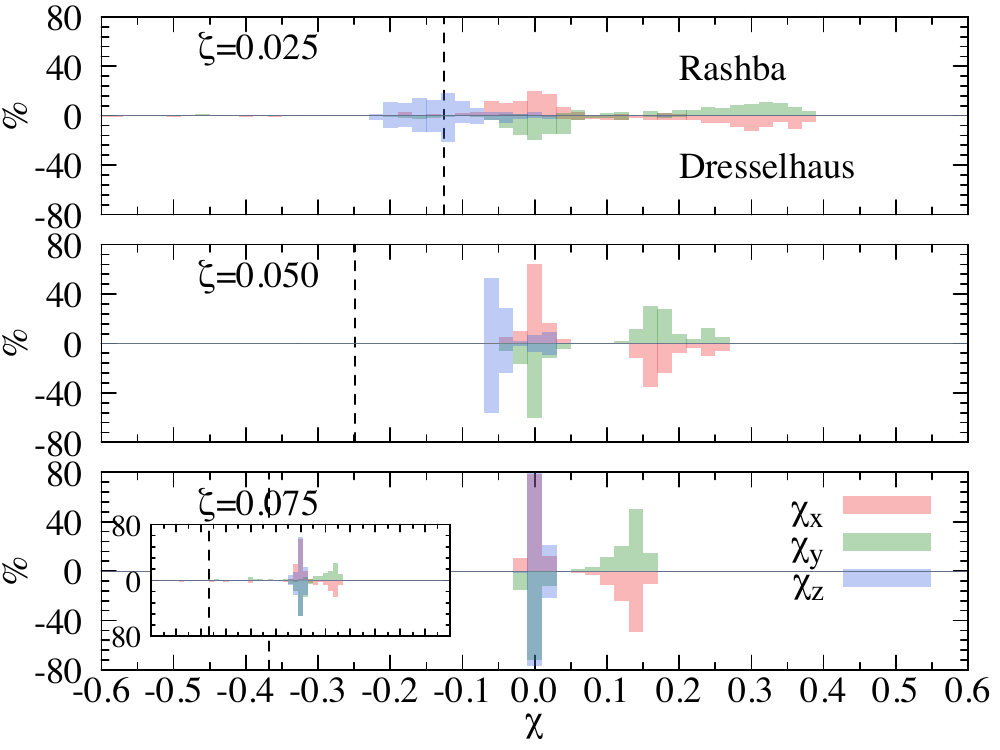}
\caption{Distribution of end chirality depending on initial Rashba-Bychkov/Dresselhaus interaction. The system parameters are the same as those used in the main text and the initial configurations are same as in Fig.\,\ref{fig:chi_pol}a. (a), (b) and (c) correspond to different magnitudes of the corresponding interactions ($\zeta$) which is shown in the legend. Inset of (c) shows the distribution with smaller damping ($\alpha=0.1$) with other parameters remaining the same as those used for (c).}
\label{fig:chi_RD}
\end{figure}

Although the end spiral states can form with any chirality, based on the initial AFM alignment they are more probable to end up in a certain configuration which is governed by the dominating component of spin mixing interaction ($\lambda$, see main text). As shown in Fig.\,3b in main text, the spiral formation starts from the edges and depending on the initial polarisation a specific component emergent chirality can dominate (Fig.~\ref{fig:chi_z}). For a fast relaxation, which can be implemented by a stronger damping parameter, the end chirality of the complete system is governed by the dominating edge component (Fig. \ref{fig:chi_pol}). For slower relaxation, the emergent chirality originating from the bulk can interfere more with the edge chirality resulting a flattening of end chirality distribution.
The end chirality of the system can be further tuned by introducing a small spin dependent hopping such as Rashba-Bychkov or Dresselhaus ($H_{R,D} = \zeta \sum_{\langle ij \rangle } c_i^\dagger t^{R,D}_{ij} c_{j} + H.C.$, where $t^{R,D}_{ij}=\pm i\sigma_{y,x}$ for $j= i \pm \hat{x}$ and $\zeta$ is the strength of interaction) type interaction which promotes a particular type of spin mixing and thus lead to a specific end configuration (Fig.~\ref{fig:chi_RD}).

\section{Thermal Excitation}

\begin{figure}[ht!]
\centering
\includegraphics[width=0.48\textwidth]{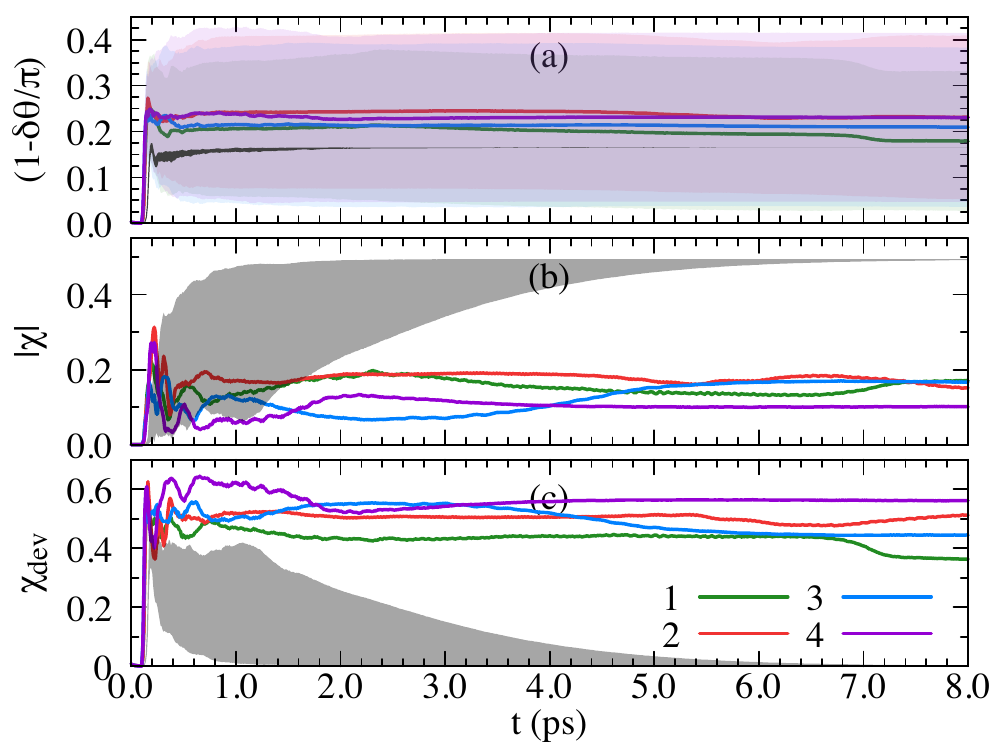}
\caption{ Evolution of  (a) relative angle ($\delta \theta$), (b) chirality ($|\chi|$) and (c) chiral deviation ($\chi_{dev}$) for four different initial configuration with thermally excite dynamics. Shaded region shows the rms deviation. Black shaded region shows the area spanned by the 64 different evolution excited by laser (Fig.\ref{fig:avg-chi}). The excitation takes place at 0.1ps.}
\label{fig:chi_th4}
\end{figure}

\begin{figure}[h!]
\centering
\includegraphics[width=0.48\textwidth]{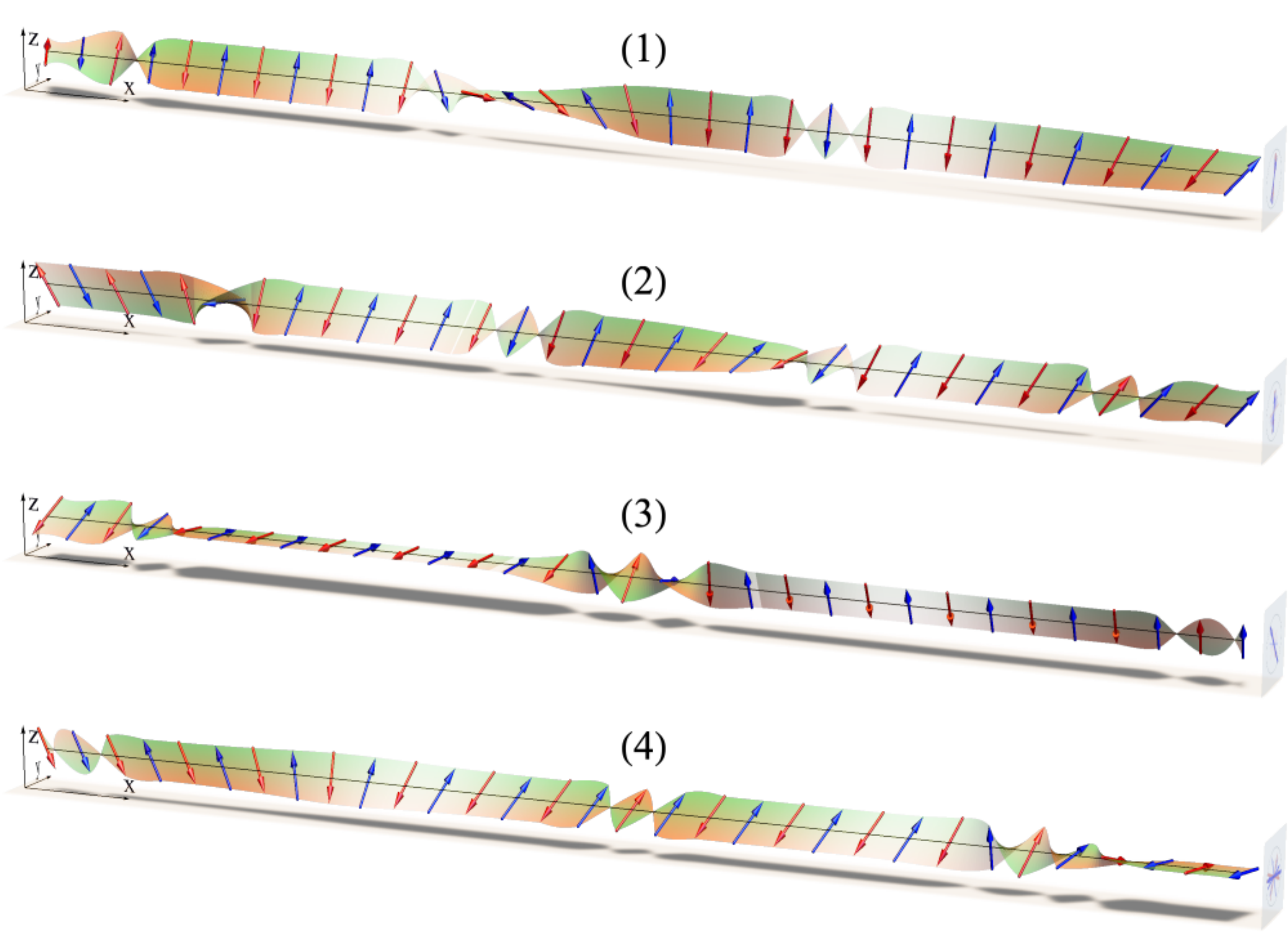}
\caption{End configuration at 8ps for four different cases shown with lines  in Fig.\,\ref{fig:chi_th4}.}
\label{fig:spiral4}
\end{figure}

Here we show time evolution of four different initial configurations for a better overview of thermally excited magnetisation dynamics. For each cases with thermal excitation the system forms multiple domains rather than a smooth spiral (Fig.~\ref{fig:spiral4}). In accord to existing knowledge, such formation can change the effective magnetic order which can be observed in experiments on ultrafast demagnetisation. However the large rms deviation of $\delta \theta$ (Fig.~\ref{fig:chi_th}a) and significantly large value of $\chi_{dev}$ (Fig.~\ref{fig:chi_th4}c) compared to laser-excited dynamics shows that the thermal excitation does not guide the system to a spiral formation. This has been further clarified by showing the end configuration of the evolution (Fig.~\ref{fig:spiral4}). Note that there are sudden jumps in different observables for certain configurations, which indicates that the end configurations are trapped in local energy minima and can spontaneously come down to a lower energy configuration. Compared to that, in laser excited dynamics, the system moves towards in a smooth and monotonic fashion denoting the coherence of the process.

\section{Effect of ambient temperature}

\begin{figure}[h]
\centering
\includegraphics[width=0.45\textwidth]{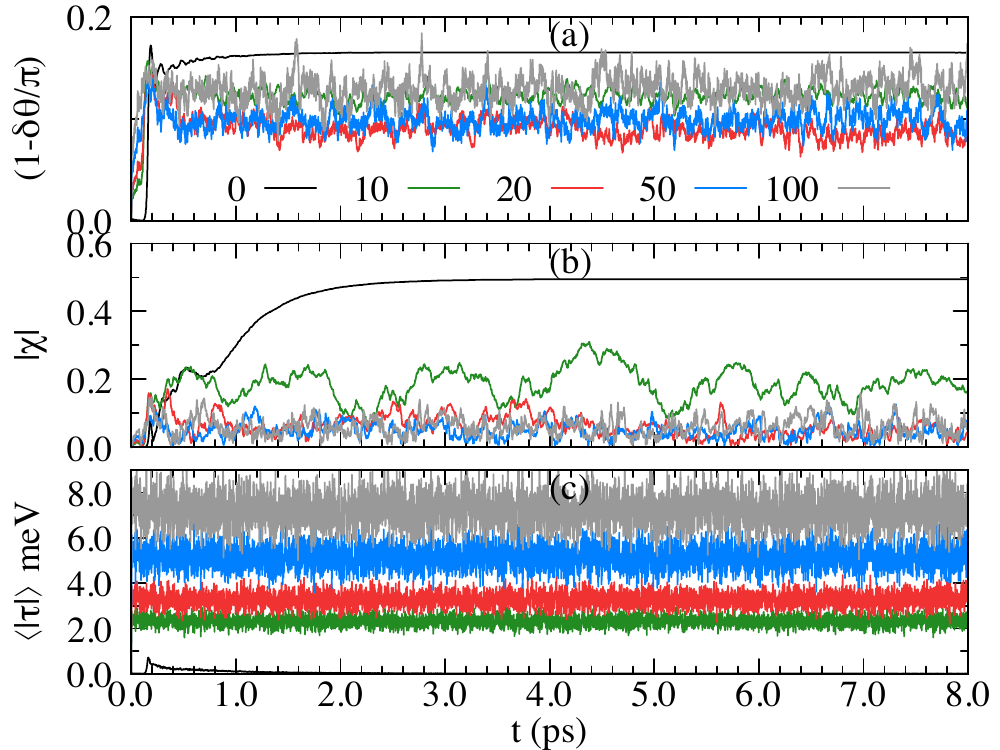}
\caption{Evolution of different entities over time for different temperatures. Black, green, red, blue and gray shows the trajectory for 0, 10, 20, 50 and 100 K for (a) angular deviation, (b) chirality and (c) average torque.}
\label{fig:thfl}
\end{figure}

As described in the main text, the thermal fluctuation is modelled as a random effective field where each component follows a normal distribution. Here we show the effect of thermal fluctuation on the evolution (Fig.\ref{fig:thfl}). We choose the same initial configuration as in Figs.\,1-2 of the  main text. 

\end{document}